\documentclass[12pt]{iopart}

\RequirePackage{graphicx}
\RequirePackage{latexsym}
\usepackage{epstopdf}
\usepackage{amssymb}
\usepackage{xcolor}

\begin{document}

\title[Particle Dynamics on Test Papapetrou Fields of Vacuum Spacetimes
]{Particle Dynamics on Test Papapetrou Fields of Vacuum Spacetimes
}

\author{Rodrigo Maier}

\address{Departamento de F\'isica Te\'orica, Instituto de F\'isica, \\
Universidade do Estado do Rio de Janeiro,\\
Rua S\~ao Francisco Xavier 524, Maracan\~a, CEP20550-900, Rio de Janeiro, Brazil}
\ead{rodrigo.maier@uerj.br}
\vspace{10pt}
\begin{indented}
\item[]October 2022
\end{indented}

\begin{abstract}
In this paper we examine the dynamics of test particles subjected to Papapetrou fields 
which emerge from vacuum spacetimes isometries.
Assuming that the background geometry is described by a Kerr metric whose timelike Killing vector satisfies Maxwell equations, we evaluate the electromagnetic fields  
-- Papapetrou fields -- by
fixing a locally non-rotating (LNR) frame of reference. 
We also evaluate such the electromagnetic fields for a vanishing rotation parameter assuming a timelike frame of reference
for a Schwarzschild background.
In order to probe for the effect of the test Papapetrou fields we study the motion of charged particles in a Kerr background.
Restricting ourselves to orbits in the equatorial plane with LNR initial conditions we show that there is an explicit deviation between orbits of neutral and charged particles in the case of repulsive electromagnetic configurations. For critical charge-mass ratios $\zeta_*$ test particles can be found in the Kerr retrograde photon sphere.
For static configurations we show that massive/charged test particles may populate the 
unstable Schwarzschild photon sphere for given domain of the parameter space.
\end{abstract}

%
%
%
%
%

\section{Introduction}

Based on mathematical studies of General Relativity 
several analogies between gravitation and electromagnetism have been developed \cite{harris,Goulart:2008ab}. One of particular interest due to A. Papapetrou \cite{Papapetrou:1966zz} infers that Killing vectors
may be regarded as vector potentials which generate electromagnetic
fields -- Papapetrou fields -- satisfying the covariant Lorentz gauge and 
Maxwell-like equations.
In this paper we examine the motion of charged test particles subjected to
Papapetrou fields of vacuum spacetimes. The motivation behind the study of such fields relies on features of significant role. 

Firstly, it is well known that Maxwell equations in curved spacetime
may present a quite involved structure of difficult resolution. In this context,
test Papapetrou fields may shed some light on regular solutions of important meaning 
-- for instance, in the context of electromagnetic perturbations of black holes \cite{chandra}. 
Although there are no procedures to obtain such solutions for a general metric, 
they can be obtained once one considers Papapetrou fields which emerge from
Killing vector fields of vacuum spacetimes. In fact, in \cite{Fayos:1999de} the authors
presented a self-contained contribution on this matter, furnishing a procedure
in which the integrability conditions of the Maxwell equations are differential equations of second order
in a Newman-Penrose basis.

Moreover, astrophysical black holes are usually embedded in a complex environment 
of plasma/magnetic fields with the presence of accretion disks. In this context,
the mechanism for the electromagnetic extraction
of energy has remained uncertain \cite{Santos:2023uka,Rueda:2023mtp}. In the case of spiraling black holes for instance,
spacetime rotation can generate electromagnetic outflows similar to the classical Faraday disk
triggering Blandford-Znajek processes \cite{Blandford:1977ds}. In \cite{Morozova:2013ina} on the other hand,
rotating black holes moving
at constant velocity in an asymptotically uniform magnetic
test field were considered and electromagnetic energy losses -- from charged particles accelerated along the magnetic field
lines -- were evaluated. A more detailed analysis of such configurations can be found in \cite{Aranha:2021zwf}
where the electromagnetic fields are based on the well known Papapetrou-Wald solution \cite{Wald:1974np}. 
In fact, it has been shown that test Papapetrou fields \cite{Wald:1974np}
are fine candidates to model an external uniform magnetic field aligned
with the Kerr black hole rotation axis originating a so-called inner
engine responsible for the emission of gamma-ray bursts \cite{Ruffini:2018aiq}.
In \cite{Moradi:2019zki} the Papapetrou-Wald solution is considered in order to
furnish an effective charge 
which originates from the gravitomagnetic interaction
of a Kerr black hole with the surrounding magnetic field \cite{Rueda:2019phh}. 
It has been argued that the
existence of this effective charge justifies the assumption that the Kerr-Newman black hole is a transient stage
to approach the analysis of quantum electrodynamical processes of rotating black holes \cite{Damour:1974qv}.

For the last but not least, it can be shown that the physical electromagnetic field
of a Kerr-Newman black hole can be generated directly from the Papapetrou field obtained from 
the timelike Killing vector of a Kerr spacetime. In this paper we explicitly show this property. In the case of a vanishing rotation, the physical electromagnetic field
of a Reissner-Nordstr\"om black hole is the same as the Papapetrou field generated from a timelike Killing vector of a Schwarzschild spacetime, as one should expect. Following a similar approach
one should expect that the Papapetrou fields generated by
boosted rotating black holes \cite{Aranha:2021zwf} could lead to boosted Kerr-Newman solutions. 
Overall, the fact that such Papapetrou fields find physical counterparts 
-- such as external electromagnetic fields generated from charged black holes --
assigns a physical meaning to such fields thus justifying its study.

In the light of the above the study of Papapetrou fields around black holes 
is well motivated in the sense that (i) it could lead to new regular 
solutions of Maxwell equations in curved spacetime; (ii)
help to better understand
the physics behind processes of black hole energy extraction; (iii) lead to new solutions of Einstein field equations which describe
physical configurations. In this paper we examine the motion of charged test particles subjected to
Papapetrou fields generated from a Kerr and a Schwarzschild black holes.
Considering circular orbits in a neighbourhood of the event horizon we show 
that massive test particles are able to populate classic black hole photon spheres \cite{Bardeen:1972fi},
thus assigning a new physical feature to Papapetrou fields.

Although the above paragraphs motivate the problem we intend to address in this paper,
it is worth to mention that there are several papers in the literature discussing Papapetrou Fields in different circumstances. In \cite{Ferrando:2003dc} for instance, the authors examine type I solutions in which the Killing $2$-form is aligned with a principal bivector of the Weyl tensor. In \cite{Ferrando:2003sn} they show that such solutions support the Kasner and Taub metrics. In \cite{Nouri-Zonoz:2014bpa} on the other hand, the 
authors introduce junction conditions on non-null hypersurfaces -- regarding the Papapetrou field as a gravitoelectromagnetic tensor -- and apply to the case
of the Van Stockum solutions\cite{van}. In spite of the fact that these topics are rather mathematical,
we intend to address them in our future research.

We organize the paper as follows. In Section \ref{sec:1} we set our conventions
and obtain Maxwell-like equations satisfied by Papapetrou fields. We also show that
the Papapetrou field from the timelike Killing vector of a Kerr spacetime generates the  
physical electromagnetic field of Kerr-Newman black hole.
In Section \ref{sec:2} we fix proper observers and evaluate the corresponding electric and magnetic fields which emerge from timelike Killing vectors of a Kerr and Schwarzschild spacetimes.
The motion of charged test particles subjected to the influence of such Papapetrou fields is examined in \ref{sec:3}.
In Section \ref{sec:4} we leave our final remarks.

\section{Papapetrou Fields}
\label{sec:1}

To start, let us consider a general Faraday tensor built from a vector field $\xi_\mu$, that is
\begin{eqnarray}
\label{fmunu}
F_{\mu\nu}=\xi_{\mu;\nu}-\xi_{\nu;\mu}.    
\end{eqnarray}
Assuming that $\xi_\mu$ is also a Killing vector we obtain
\begin{eqnarray}
\label{fmunu2}
F_{\mu\nu}=2\xi_{\mu;\nu}    
\end{eqnarray}
so that
\begin{eqnarray}
\label{max}
F^{\mu\nu}_{~~;\nu}=2\Box\xi^\mu. 
\end{eqnarray}
However, from the integrability condition
\begin{eqnarray}
\label{ricci1}
\xi_{\mu;\nu;\gamma}-\xi_{\mu;\gamma;\nu}=-R^\sigma_{~\mu\nu\gamma}\xi_\sigma,    
\end{eqnarray}
we obtain
\begin{eqnarray}
\label{eqper}
\xi_{\gamma;\mu;\nu}= R^\sigma_{~\nu\gamma\mu}\xi_\sigma,    
\end{eqnarray}
where we have used cyclic permutations of (\ref{ricci1}) together with the Killing equation $\xi_{\gamma;\nu}+\xi_{\nu;\gamma}=0$. Therefore, from (\ref{eqper}) we end up with
\begin{eqnarray}
\label{eqsource}
\Box \xi_\gamma= R^{~\sigma}_{\gamma} \xi_\sigma,
\end{eqnarray}
and the substitution of  (\ref{eqsource}) in (\ref{max}) furnishes
\begin{eqnarray}
\label{max2}
F^{\mu\nu}_{~~;\nu}=2R^{\mu\sigma}\xi_\sigma.  
\end{eqnarray}
From the above we see that given a general metric, a Faraday tensor may be built with its Killing vectors so that Maxwell-like equations 
of the type (\ref{max2}) must be satisfied. The source term $J^\mu\propto R^{\mu\sigma}\xi_\sigma$ plays the role
of a $4$-current given by the projection of the Ricci tensor in the direction of the corresponding Killing vector.  
In this context the electromagnetic fields which arise from $F_{\mu\nu}$ -- as their source terms -- rely on the same geometrical feature, namely, spacetime isometries. We refer to such fields as Papapetrou fields \cite{Papapetrou:1966zz,Fayos:1999de}.

Although Papapetrou fields are usually taken as test electromagnetic fields,
they present a physical signature when vacuum spacetimes are considered.
In fact, it can be shown that the Papapetrou field which emerge from the timelike Killing vector of
a Kerr black hole is actually connected to physical electromagnetic fields of Kerr-Newman black holes \cite{Wald:1974np,misner}. 
To explicitly show this behaviour, 
let us consider 
a Kerr spacetime in Kerr-Schild coordinates:
\begin{eqnarray}
\label{kerr0}
\nonumber
&&ds^2=-\Big[1-\frac{2mr}{\Sigma(r, \theta)}\Big]dt^2+\frac{4mr}{\Sigma(r, \theta)}dt dr- \frac{4mra \sin^2{\theta}}{\Sigma(r, \theta)}dt d\phi
\\
&&~~~~~~~~~~~~+\Big[1+\frac{2mr}{\Sigma(r, \theta)}\Big]dr^2-2a\sin^2{\theta}\Big[1+\frac{2mr}{\Sigma(r, \theta)}\Big]dr d\phi\\
\nonumber
&&~~~~~~~~~~~~~~~~~~~~~~~~~~+\Sigma(r, \theta)d\theta^2+\frac{A(r, \theta)}{\Sigma(r,\theta)}\sin^2{\theta}d\phi^2,
\end{eqnarray}
where $m$ and $a$ are the geometrical mass and angular momentum, respectively. Furthermore,
\begin{eqnarray}
\label{sigma}
\Sigma(r, \theta)&=&r^2+a^2\cos^2\theta,\\
\label{A}
A(r,\theta)&=&\Sigma(r, \theta)\Delta(r)+2mr(r^2+a^2),\\
\label{delta}
\Delta(r)&=&r^2-2mr+a^2.
\end{eqnarray}
For a given constant parameter $\epsilon_0$ it is easy to show that 
\begin{eqnarray}
\label{sch2}
K^\mu=(\epsilon_0,0,0,0),    
\end{eqnarray}
is a timelike Killing vector of Kerr spacetime. Assuming that $K^\mu$ sources the Maxwell field in the sense that (\ref{eqsource}) is satisfied 
for the Kerr metric (\ref{kerr0}), we notice that the nonvanishing components of the energy momentum-tensor 
\begin{eqnarray}
\label{tmaxwell}
T^{\mu}_{~\nu}=F^{\mu\alpha}F_{\nu\alpha}-\frac{1}{4}F^{\alpha\beta}F_{\alpha\beta}\delta^\mu_{~\nu} \end{eqnarray}
constructed from (\ref{kerr0})-(\ref{sch2}) -- with the use of (\ref{fmunu2}) -- read
\begin{eqnarray}
\nonumber
&&T^{t}_{~t}=-T^{\phi}_{~\phi}=-\frac{\epsilon^2_0 m^2}{2}\Big[\frac{r^2+a^2(1+\sin^2\theta)}{(r^2+a^2\cos^2\theta)^3}\Big],\\
\nonumber
&&T^{t}_{~r}=-\frac{\epsilon^2_0 m^2a^2\sin^2\theta}{(r^2+a^2\cos^2\theta)^3},\\
\nonumber
&&T^{t}_{~\phi}=\frac{\epsilon^2_0 m^2a\sin^2\theta(r^2+a^2)}{(r^2+a^2\cos^2\theta)^3},\\
\nonumber
&&T^{r}_{~r}=-T^{\theta}_{~\theta}=-\frac{\epsilon^2_0 m^2}{2(r^2+a^2\cos^2\theta)^2},\\
\nonumber
&&T^{\phi}_{~t}=T^{\phi}_{~r}=-\frac{\epsilon_0m^2a}{(r^2+a^2\cos^2\theta)^3}.
\end{eqnarray}
It turns out that these components of the energy-momentum tensor satisfy Einstein-Maxwell equations for the Kerr-Newman metric with mass $m$ and charge $q$
\begin{eqnarray}
\nonumber
ds^2=-\Big(1-\frac{2mr-{q}^2}{\Sigma}\Big)dt^2+\frac{2(2mr-{q}^2)}{\Sigma}dtdr\\
\nonumber
~~~~~~~~~~-\frac{2a(2mr-{q}^2)}{\Sigma}\sin^2\theta dt d\phi+\Big(1+\frac{2mr-{q}^2}{\Sigma}\Big)dr^2\\
~~~~~~~~~~-2a\sin^2\theta\Big(1+\frac{2mr-{q}^2}{\Sigma}\Big)drd\phi+\Sigma d\theta^2
\\
\nonumber
~~~~~~~~~~+\frac{\sin^2\theta}{\Sigma}[(r^2+a^2)^2-\tilde{\Delta}a^2\sin^2\theta]d\phi^2,
\end{eqnarray}
where
\begin{eqnarray}
\tilde{\Delta}(r)=r^2-2mr+(a^2+{q}^2)    
\end{eqnarray}
and
\begin{eqnarray}
\label{eqe0}
\epsilon_0=\frac{\sqrt{2}{q}}{m}.    
\end{eqnarray}
From (\ref{eqe0}) we see how the electric charge $q$ of a Kerr-Newman black hole can be written in terms of $\epsilon_0$ -- which describes the timelike killing vector of a Kerr black hole. 
In the case of zero rotation,
the Papapetrou field of the timelike Killing vector of
a Schwarzschild black hole is connected to physical electromagnetic field of a Reissner-Nordstr\"om
black hole, as one should expect.

In this paper we intend to examine the effect of test Papapetrou fields on particle motions. To this end we consider a general Kerr spacetime in Kerr-Schild coordinates (\ref{kerr0}) and fix a proper locally non-rotating (LNR) frame of reference \cite{Takahashi:2007yi}. An equivalent analysis is performed for the Schwarzschild spacetime
for a suitable timelike observer.

\section{Papapetrou Fields in Vacuum Spacetimes}
\label{sec:2}

We start by considering 
a Kerr spacetime in Kerr-Schild coordinates (\ref{kerr0}).
Let
\begin{eqnarray}
\label{k2}
\xi^\mu=(\epsilon_0, 0, 0, \gamma_0),    
\end{eqnarray}
with $\gamma_0$ also constant, be a general Killing vector of Kerr spacetime.
Therefore, 
the nonvanishing components of the Faraday tensor $F_{\mu\nu}$ read
\begin{eqnarray}
\label{f2}
&&F_{tr}=2m(\epsilon_0-\gamma_0 a\sin^2\theta)\frac{d}{dr}\Big(\frac{r}{\Sigma}\Big),\\
&&F_{t\theta}=2mr\Big[\epsilon_0 \frac{d}{d\theta}\Big(\frac{1}{\Sigma}\Big)-\gamma_0 a\frac{d}{d\theta}\Big(\frac{\sin^2\theta}{\Sigma}\Big)\Big],\\
&&F_{r\theta}=2mr\epsilon_0\frac{d}{d\theta}\Big(\frac{1}{\Sigma}\Big)-\gamma_0 a \frac{d}{d\theta}\Big[\sin^2\theta\Big(1+\frac{2mr}{\Sigma}\Big)\Big],~~~~~~\\
&&F_{r\phi}=\sin^2\theta\frac{d}{dr}\Big(\frac{2mar \epsilon_0-\gamma_0A}{\Sigma}\Big),\\
\label{f6}
&&F_{\theta\phi}=\frac{d}{d\theta}\Big[\sin^2\theta\Big(\frac{2mar\epsilon_0-\gamma_0A}{\Sigma}\Big)\Big].
\end{eqnarray}
Defining the Hodge dual by 
${{\cal F}}^{\mu\nu}=\frac{1}{2}\epsilon^{\mu\nu\alpha\beta} {F}_{\alpha\beta}$ we
obtain 
\begin{eqnarray}
\nonumber
&&{\cal F}^{tr}=  F_{\theta\phi},~~{\cal F}^{t\theta}= - F_{r\phi},~~
{\cal F}^{t\phi}= F_{r \theta},~~{\cal F}^{r\theta}=0,\\
\nonumber
&&{\cal F}^{r\phi}= - F_{t\theta},~~{\cal F}^{\theta\phi}= F_{tr}.  
\end{eqnarray}
\begin{figure}
\begin{center}
\includegraphics[width=6cm,height=6cm]{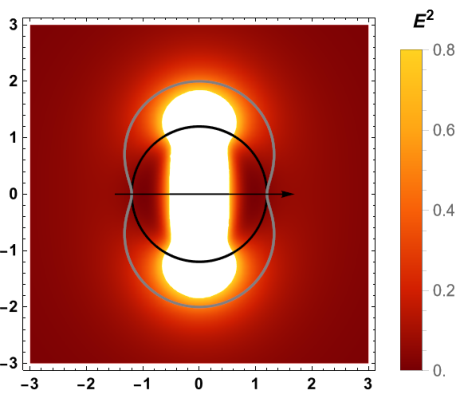}
\includegraphics[width=6cm,height=6cm]{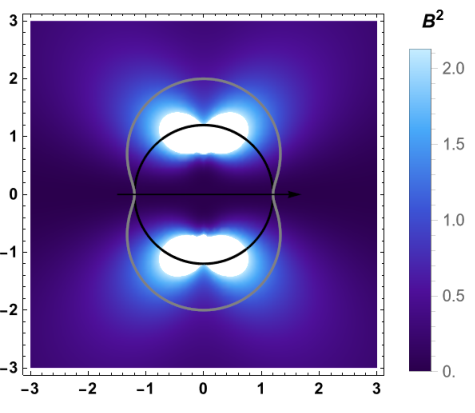}
\caption{The magnitude of the electric (left panel) and magnetic (right panel) fields in the section $\sin\phi=0$. 
Here we have fixed the parameters $m=1.00$, $a=0.98$ and $\epsilon_0=1.00$.
The event horizon $r_+$ and ergosphere $r_e=m+\sqrt{m^2-a^2\cos^2\theta}$ are illustrated by the black and grey solid curves, respectively.
The black solid arrow indicates the axis of rotation of the black hole.}
\label{fig1}
\end{center}
\end{figure}

In order to evaluate the electromagnetic fields we now fix a proper locally non-rotating (LNR) reference frame \cite{Takahashi:2007yi} whose observer's $4$-velocity $u^\mu$ is given by
\begin{eqnarray}
u^\mu=\frac{1}{N}(1, -N^r, 0, 0),    
\end{eqnarray}
where 
\begin{eqnarray}
\label{lapshift}
N=\Big(1+\frac{2mr}{\Sigma}\Big)^{-\frac{1}{2}},~~N^r=\frac{2mr}{\Sigma+2mr},    
\end{eqnarray}
are the lapse and the shift functions, respectively, according to the ADM decomposition \cite{Arnowitt:1959ah} of (\ref{kerr0}). In this case, 
the components of the electric 
and magnetic fields for the Killing vector (\ref{k2}) are evaluated by 
\begin{eqnarray}
E^\mu=F_{\mu\nu}u^\nu,    
\end{eqnarray}
and
\begin{eqnarray}
B^\mu = {\cal F}^{\mu}_{~\nu} u^\nu,     
\end{eqnarray}
respectively.

However, a straightforward evaluation of the magnitude of the magnetic field in the equatorial plane ($\theta=\pi/2$)
furnishes
\begin{eqnarray}
\vec{B}^2=\frac{4[\epsilon_0 ma+\gamma_0(r^3-ma^2)]^2}{r(2m+r)},
\end{eqnarray}
implying that $\vec{B}^2$ diverges as $r\rightarrow+\infty$ for a nonvanishing $\gamma_0$.
This result assigns a rather mathematical meaning -- with no physical grounds -- for the electromagnetic field which arises from the axial component of the Killing vector.
For this reason, contrary to the analysis performed in \cite{Lee:2022rtg}, in the following we shall restrict ourselves to
pure timelike Killing vectors. In this case, the components of the electric field constructed from (\ref{f2})-(\ref{f6}) with
$\gamma_0\equiv 0$ read
\begin{eqnarray}
\label{er}
&&E^r=\frac{8m\epsilon_0(r^2+a^2)(r^2-a^2\cos^2\theta)}{\sqrt{1+\frac{2mr}{r^2+a^2\cos^2\theta}}(r^2+a^2\cos^2\theta)^3},\\
&&E^\theta=-\frac{2m\epsilon_0a^2r\sin(2\theta)}{\sqrt{1+\frac{2mr}{r^2+a^2\cos^2\theta}}(r^2+a^2\cos^2\theta)^3},\\
&&E^\phi=\frac{8m\epsilon_0 a(r^2-a^2\cos^2\theta)}{\sqrt{1+\frac{2mr}{r^2+a^2\cos^2\theta}}(r^2+a^2\cos^2\theta)^3},
\end{eqnarray}
while the magnetic counterparts are given by 
\begin{eqnarray}
&&B^r=\frac{2m\epsilon_0ar(r^2+a^2)\sin{(2\theta)}}{\sqrt{1+\frac{2mr}{r^2+a^2\cos^2\theta}}(r^2+a^2\cos^2\theta)^2},\\
&&B^\theta=\frac{2m\epsilon_0a(r^2-a^2\cos^2\theta)\sin^2\theta}{\sqrt{1+\frac{2mr}{r^2+a^2\cos^2\theta}}(r^2+a^2\cos^2\theta)^2},\\
\label{bphi}
&&B^\phi=\frac{2m\epsilon_0 a^2r\sin(2\theta)}{\sqrt{1+\frac{2mr}{r^2+a^2\cos^2\theta}}(r^2+a^2\cos^2\theta)^2}.
\end{eqnarray}
Finally, the magnitude of the electric and magnetic fields read:
\begin{eqnarray}
\nonumber
\vec{E}^2=\frac{4m^2\epsilon_0^2}{(r^2+a^2\cos^2\theta)^4}\Big\{\frac{(r^2+a^2)(r^2-a^2\cos^2\theta)^2}{(r^2+a^2\cos^2\theta)}\\
\label{efp}
+\frac{a^4r^2\sin^2(2\theta)}{r(2m+r)+a^2\cos^2\theta}-\frac{2ma^2r(r^2-a^2\cos^2\theta)^2\sin^2\theta}{(r^2+a^2\cos^2\theta)[r(2m+r)+a^2\cos^2\theta]}\Big\},\\
\nonumber
\vec{B}^2=\frac{4m^2\epsilon_0^2a^2}{(r^2+a^2\cos^2\theta)^2}\Big\{\frac{r^2(r^2+a^2)\sin^2(2\theta)}{(r^2+a^2\cos^2\theta)}+\frac{(r^2-a^2\cos^2\theta)^2\sin^4\theta}{r(2m+r)+a^2\cos^2\theta}\\
\label{bfp}
~~~~~~~~-\frac{8m a^2r^3\cos^2\theta\sin^4\theta}{(r^2+a^2\cos^2\theta)[r(2m+r)+a^2\cos^2\theta]}\Big\}.
\end{eqnarray}

In Kerr-Schild coordinates, the event horizon $r_+$ and ergosphere $r_e$ are given by
\begin{eqnarray}
\nonumber
r_+=m+\sqrt{m^2-a^2},~~r_e=m+\sqrt{m^2-a^2\cos^2\theta}.    
\end{eqnarray}
It can be shown that the electric field has a high magnitude domain in the region between the event horizon
and the ergosphere in a finite neighbourhood of $\theta=\pi/2$. The magnetic field, on the other hand, has a high magnitude domain in the region between the event horizon
and the ergosphere in a finite neighbourhood of $\theta=\pi/4$.
In order to illustrate such a behaviour, let us consider the parameters $m=1.00$, $a=0.98$ and $\epsilon_0=1.00$.
In Fig. 1 we show a numerical plot of the magnitude of the electric and magnetic fields in a neighbourhood of the event horizon and ergosphere.  It is worth to mention that the analytical solutions (\ref{efp})--(\ref{bfp}) have the desirable
asymptotic behavior once
\begin{eqnarray}
\lim_{r\rightarrow +\infty}\vec{E}^2=\frac{4m^2\epsilon_0^2}{r^4}=\frac{8q^2}{r^4}~~{\rm and}~~\lim_{r\rightarrow +\infty}\vec{B}^2=0.
\end{eqnarray}
In the above we have used (\ref{eqe0}) so that asymptotically
the electromagnetic field of the rotating black hole reduces
to the Coulomb potential showing the electromagnetic nature
of the Papapetrou field.

Although the above results can be applied to the case of a Schwarzschild spacetime for $a=0$,
a more suitable frame of reference can be chosen once we write the Schwarzschild metric
in the another coordinate system such that
\begin{eqnarray}
\label{sch1}
ds^2=-\Big(1-\frac{2m}{r}\Big)dt^2+\Big(1-\frac{2m}{r}\Big)^{-1}dr^2
+r^2d\theta^2+r^2\sin^2\theta d\phi^2.
\end{eqnarray}
To differ from the Kerr case, we chose another constant parameter $s_0$ so that 
\begin{eqnarray}
\label{sch3}
S^\mu=(s_0,0,0,0),    
\end{eqnarray}
is a timelike Killing vector of Schwarzschild spacetime (\ref{sch1}).
Regarding (\ref{sch3}) as a test electromagnetic potential it is then easy to show from (\ref{fmunu2}) that the sole nonvanishing independent component of the Faraday tensor read
\begin{eqnarray}
F_{tr}=-\frac{2ms_0}{r^2},
\end{eqnarray}
and, by definition, ${\cal F}^{\theta\phi}=F_{tr}$.

In order to evaluate the electromagnetic fields we now fix a proper timelike reference frame whose observer's $4$-velocity $u^\mu$ is given by
\begin{eqnarray}
\label{fch}
u^\mu=\delta^{\mu}_{~t}.   
\end{eqnarray}
A straightforward evaluation shows that the components of the magnetic field vanish 
while the electric counterparts read
\begin{eqnarray}
\label{ersch}
E^r=\frac{2ms_0}{r^2}\Big(1-\frac{2m}{r}\Big),~~E^\theta=E^\phi\equiv 0.
\end{eqnarray}
The magnitude of the electric field can then be easily evaluated furnishing:
\begin{eqnarray}
\vec{E}^2=\Big(\frac{2ms_0}{r^2}\Big)^2\Big(1-\frac{2m}{r}\Big).
\end{eqnarray}
It can be analytically shown that the magnitude of the electric field is well behaved in the
domain $r\geq 2m$. In fact, $\vec{E}^2$ has a global maximum 
at $r_{max}=5m/2$ and vanishes for $r=2m$.
As $r\rightarrow+\infty$, $\vec{E}^2\propto r^{-4}$ reflecting the Coulomb potential as one should expect.
The radial component of the electric field, on the other hand, exhibits a more interesting behaviour
in the reference frame (\ref{fch}).
In fact it can be shown that it is also well behaved in the
domain $r\geq 2m$. However, it reaches its global maximum 
exactly at the Schwarzschild photon shere located at $r=3m$ and vanishes for $r=2m$.
In Fig. 2 we show a numerical plot of the radial component of the electric field in a neighbourhood of the event horizon.
\begin{figure}
\begin{center}
\includegraphics[width=6cm,height=4cm]{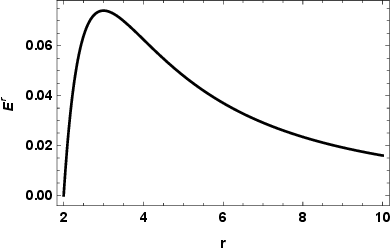}
\caption{The radial component of the electric field in a neighbourhood of the event horizon. 
Here we have fixed the parameters $m=s_0=1.00$.
In the above we see that $E^r$ is well behaved in a neighbourhood of the event horizon
and reaches its maximum over the Schwarzschild photon sphere located at $r=3m$.}
\label{fig0}
\end{center}
\end{figure}

\section{Particle Dynamics}
\label{sec:3}

To proceed we now investigate the motion of charged test particles subjected to Papapetrou fields of vacuum spacetimes.
A particle moving in a gravitational and electromagnetic field has equations of motion
\begin{eqnarray}
\label{geo}
\frac{d^2x^\mu}{d\tau^2}
+\Gamma^\mu_{~\alpha\beta}
\frac{dx^\alpha}{d\tau}\frac{dx^\beta}{d\tau}=\zeta F^\mu_{~\alpha}\frac{dx^\alpha}{d\tau},   
\end{eqnarray}
where we regard $\Gamma^\mu_{~\alpha\beta}$ as the Christoffel connection, $\tau$ is the proper time and $\zeta$ is the charge-mass ratio of the particle, namely, $\zeta\equiv \tilde{q}/\tilde{m}$. It can be easily shown that the equations of motion (\ref{geo}) can be obtained from the Lagrangian
\begin{eqnarray}
\label{lag}
{\cal L}=\frac{1}{2}g_{\mu\nu}\frac{dx^\mu}{d\sigma}\frac{dx^\nu}{d\sigma}-\tilde{q} A_\mu \frac{dx^\mu}{d\sigma}.     
\end{eqnarray}
In the above we define the affine parameter as $\sigma=\tau/\tilde{m}$. 

In order to furnish first integrals of the equations of motion (\ref{geo}) we now turn 
to the hamiltonian formulation. From (\ref{lag}) we obtain that the canonical momenta are given by
\begin{eqnarray}
\label{pmu}
p^\mu=\frac{dx^\mu}{d\sigma}-\tilde{q}A^\mu,     
\end{eqnarray}
so that the Hamiltonian reads
\begin{eqnarray}
\label{ham2}
H=\frac{1}{2}g_{\mu\nu}({p^\mu}+\tilde{q}A^\mu)({p^\nu}+\tilde{q}A^\nu)=-\frac{\tilde{m}^2}{2}.    
\end{eqnarray}
\subsection{Kerr Spacetime}

In the case of a Kerr spacetime (\ref{kerr0}) together with $A^\mu=K^\mu$, $H$ does not depend on $t$ nor $\phi$. Therefore, from the above we obtain two fundamental constants of motion,
\begin{eqnarray}
\label{c1}
p_t=-E,~~p_\phi=L,    
\end{eqnarray}
connected to the energy and angular momentum, respectively. A third constant of motion can be obtained
once we employ the simple method analogous to that of Carter \cite{Carter:1968rr}. In fact, defining $S$ as the Jacobi action
so that $p_r=\partial S/\partial r$ and $p_\theta=\partial S/\partial \theta$, the Hamilton constraint
(\ref{ham2}) can be written as
\begin{eqnarray}
\nonumber
\Big(\frac{\partial S}{\partial \theta}\Big)^2+a^2\cos^2\theta[\tilde{m}^2-(E+\tilde{q}\epsilon_0)^2]+L^2\cot^2\theta~~~~\\
\label{jacobi}
+\Big(\frac{\partial S}{\partial r}\Big)^2\Delta
+2(aL-2Emr)\Big(\frac{\partial S}{\partial r}\Big)+L^2~~~~~~~~~\\
\nonumber
+r^2[\tilde{m}^2-(E+\tilde{q}\epsilon_0)^2]-2mr(E^2-\tilde{q}\epsilon_0^2)=0.    
\end{eqnarray}
From the above we define the constant of motion 
\begin{eqnarray}
\label{c2}
Q=p_\theta^2+a^2\cos^2\theta[\tilde{m}^2-(E+\tilde{q}\epsilon_0)^2]+L^2\cot^2\theta,    
\end{eqnarray}
so that the radial component satisfies
\begin{eqnarray}
\label{pr}
p_r=\frac{2Emr-aL\pm\sqrt{\Phi(r)}}{\Delta},    
\end{eqnarray}
where 
%
%
\begin{eqnarray}
\nonumber
\Phi(r)\equiv(aL-2Emr)^2-\Delta\{L^2+Q~~~~~~~~~~~~~~~~\\
-2mr(E^2-\tilde{q}^2\epsilon_0^2)-r^2[(E+\tilde{q}\epsilon_0)^2-\tilde{m}^2]\}.   
\end{eqnarray}
%
%
%

From (\ref{pmu})--(\ref{c1}), (\ref{c2}) and (\ref{pr}) we then obtain:
\begin{eqnarray}
\label{eqr1}
&&\frac{dt}{d\sigma}= E\Big(1+\frac{2mr}{\Sigma}\Big)+\frac{2mr(2Emr-aL\pm\sqrt{\Phi})}{\Sigma\Delta}+\tilde{q}\epsilon_0,
\\
&&\frac{dr}{d\sigma}=\pm\frac{\sqrt{\Phi(r)}}{\Sigma},
\\
\label{theta}
&&\frac{d\theta}{d\sigma}=\pm\frac{\sqrt{\Theta(r,\theta)}}{\Sigma},
\\
\label{eqphi1k}
&&\frac{d\phi}{d\sigma}=\frac{L\csc^2\theta}{\Sigma}+\frac{a(2Emr-aL\pm\sqrt{\Phi(r)})}{\Sigma\Delta},
\end{eqnarray}
where
\begin{eqnarray}
\nonumber
\Theta(r,\theta)=Q+a^2[(E+\tilde{q}\epsilon_0)^2-m^2]\cos^2\theta-L^2\cot^2\theta.
\end{eqnarray}
From the above we see that physical orbits are those which satisfy the necessary condition $\Phi\geq 0$ and $\Theta\geq 0$. The former condition can be simplified 
in the domain $r\geq r_+$ by 
\begin{eqnarray}
L^2+Q-2mr(E^2-\tilde{q}^2\epsilon_0^2)-r^2[(E+\tilde{q}\epsilon_0)^2-\tilde{m}^2]\leq 0.   
\end{eqnarray}

In order to simply our analysis we restrict ourselves orbits with initial conditions $\theta_0=\pi/2$ and
$d\theta/d\sigma|_0=0$.
From (\ref{theta}) we then notice that the dynamics is restricted in the equatorial plane once $Q=0$.
Trajectories of particular interest are circular orbits with $r=r_c$ defined by
\begin{eqnarray}
\label{phi1}
&&\Phi(r)=0,\\
\label{phi2}
&&\frac{d\Phi}{dr}=0.    
\end{eqnarray}
It can be easily shown that the solution of (\ref{phi2}) is given by
\begin{eqnarray}
\label{rcsk}
r_{c\pm}=\frac{2m(2E\tilde{q}\epsilon_0-\tilde{m}^2)\pm\sqrt{\Gamma}}{3[E(E+2\tilde{q}\epsilon_0)-\tilde{m}^2]}    
\end{eqnarray}
where
\begin{eqnarray}
\nonumber
\Gamma=4m^2(2E\tilde{q}\epsilon_0-\tilde{m}^2)^2-3(E^2+2E\tilde{q}\epsilon_0-\tilde{m}^2)\\
\times[a^2(E^2+2E\tilde{q}\epsilon_0-\tilde{m}^2)-L^2].    
\end{eqnarray}
Needless to say that the substitution of (\ref{rcsk}) in (\ref{phi1}) 
is a rather involved task in order 
to find for circular orbits.
Following an alternative route to probe for circular orbits, let us consider
particular configurations in which second order terms as $\tilde{q}^2\epsilon_0^2$
may be neglected. Since we regard the Papapetrou field as a test electromagnetic field
this is a reasonable assumption which one may consider from start.
%
%
%
In this approximation
equations (\ref{phi1}) and (\ref{phi2}) can be simultaneously solved furnishing
\begin{eqnarray}
\label{lc1}
L\simeq\pm(r^{2}\mp2am^{1/2}r^{1/2}+a^{2})\sqrt{\frac{ m(\tilde{m}^2-2E\tilde{q}\epsilon_0)}{r^{3/2}(r^{3/2}-3mr^{1/2}\pm 2am^{1/2})}},
\end{eqnarray}
where $E$ satisfies
\begin{eqnarray}
\label{ec1}
E\simeq(r^{3/2}-2mr^{1/2}\pm am^{1/2})\sqrt{\frac{\tilde{m}^2-2E\tilde{q}\epsilon_0}{r^{3/2}(r^{3/2}-3mr^{1/2}\pm 2am^{1/2})}}.
\end{eqnarray}
As in the case of a null Papapetrou field, the upper sign in (\ref{lc1}) and (\ref{ec1}) 
is connected to corotating orbits \cite{Bardeen:1972fi} with $L>0$. The lower sign on the other hand refers to retrograde 
orbits with $L<0$. Therefore, circular orbits can be found as long as
\begin{eqnarray}
\label{inq}
\frac{\tilde{m}^2-2E\tilde{q}\epsilon_0}{r^{3/2}-3mr^{1/2}\pm 2am^{1/2}}    \gtrsim 0.   
\end{eqnarray}
In the case that the LRS of (\ref{inq})
is sufficiently small,
we obtain orbits in which
\begin{eqnarray}
\tilde{m}^2\simeq2E\tilde{q}\epsilon_0. 
\end{eqnarray}
Assuming that the denominator of (\ref{inq}) is also sufficiently small, one can obtain
circular trajectories analogous to photon orbits of a Kerr black hole \cite{Bardeen:1972fi} 
(similar results for charged black holes can be found in \cite{Carter:1968rr,Khan:2020sya,Khan:2020zkl}).
In the present context however, massive particles with mass
$\tilde{m}\simeq (2E \tilde{q}\epsilon_0)^{1/2}$ are able to populate such circular orbits
located at
\begin{eqnarray}
r_{ph}\simeq 2m\Big\{1+\cos\Big[\frac{2}{3}\arccos{\Big(\mp\frac{a}{m}\Big)}\Big]\Big\},    
\end{eqnarray}
arbitrarily close to the Kerr photon sphere \cite{Bardeen:1972fi}.

In order to illustrate the results discussed above, let us fix proper initial conditions in the equatorial plane connected to a LNR frame -- the same frame of reference in which the electromagnetic fields (\ref{er})--(\ref{bphi}) were evaluated. Namely,
\begin{eqnarray}
\frac{dt}{d\tau}\Big|_0=\frac{1}{N_0},~~\frac{dr}{d\tau}\Big|_0=-\frac{N^r}{N}\Big|_0,~~\frac{d\phi}{d\tau}\Big|_0=0. 
\end{eqnarray}
The remaining initial conditions are fixed with the assumption that the dynamics
evolve in a neighbourhood of the event horizon. For the case of $m=1.00$, $a=0.98$ and $\epsilon_0=1.00$, it can be shown that such assumption is fulfilled for a wide range of initial conditions in a neighbourhood of $t_0=0$, $r_0=10$ and $\phi_0=0.5$ radians. However, instead of (\ref{eqr1})--(\ref{eqphi1k}), the dynamical evolution with such initial conditions require the second order
equations of motion (\ref{geo}). Taking into account (\ref{kerr0}) and (\ref{f2})--(\ref{f6}), these equations reduce to
\begin{figure*}
\begin{center}
\includegraphics[width=7cm,height=5cm]{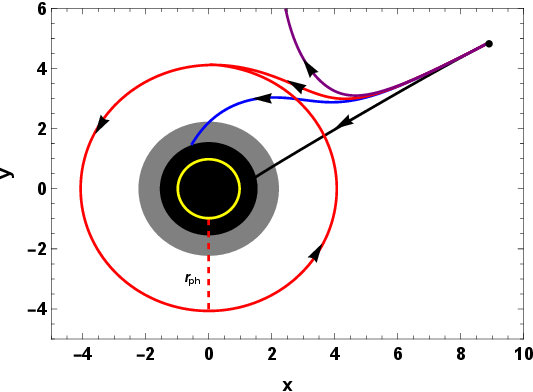}
\includegraphics[width=7cm,height=5cm]{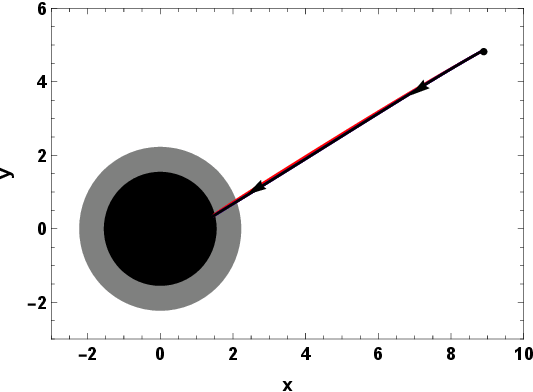}
\caption{Orbits of test particles with LNR initial conditions. The external edge of the black and grey disks stand for the event horizon and ergosphere, respectively.
For repulsive electromagnetic configurations (left panel) the charged-mass ratios read: $\zeta=0$ (neutral particles), $\zeta=0.66$, $\zeta_*\simeq0.67$ and $\zeta=0.68$, corresponding to the black, blue, red and purple orbits, respectively. 
In this case there is an explicit deviation between orbits of neutral and charged particles due to Papapetrou fields.
Charged particles with $\zeta=\zeta_*$ (red orbit)
are trapped in the unstable Kerr retrograde photon sphere -- at a distance $r\simeq r_{ph}$ from the Kerr ring singularity (yellow circle) defined by $r=0$ and $\theta=\pi/2$. 
For attractive configurations (right panel) the charge-mass ratios range from
$\zeta=0$ down to $\zeta=-0.68$ and all particles follow same trajectories towards the event horizon.}
\end{center}
\end{figure*}

\begin{eqnarray}
\nonumber
\frac{d^2t}{d\tau^2}+2m\Big[\frac{1}{r^2}\Big(1+\frac{m}{r}\Big)\Big(\frac{dr}{d\tau}\Big)^2+\frac{m}{r^3}\Big(\frac{dt}{d\tau}\Big)\frac{d}{d\tau}(t-2a\phi)\\
\label{eom1}
+\frac{1}{r^2}\Big(\frac{dr}{d\tau}\Big)\Big(1+\frac{2m}{r}\Big)\frac{d}{d\tau}(t-a\phi)
-\Big(1-\frac{ma^2}{r^3}\Big)\Big(\frac{d\phi}{d\tau}\Big)^2\Big]\\
\nonumber
=\frac{2m\epsilon_0\zeta}{r^2}\Big[\Big(1+\frac{2m}{r}\Big)\Big(\frac{dr}{d\tau}\Big)+\frac{2m}{r}\frac{d}{d\tau}(t-a\phi)\Big],\\
\nonumber
\frac{d^2r}{d\tau^2}+\frac{1}{r^4}\Big\{m\Big[a^2-r(r+2m)\Big]\Big(\frac{dr}{d\tau}\Big)^2
+2\Big(\frac{dr}{d\tau}\Big)\Big[m(a^2-2mr)\Big(\frac{dt}{d\tau}\Big)\\
-a(ma^2-2m^2r-r^3)\Big(\frac{d\phi}{d\tau}\Big)\Big]
+(a^2-2mr+r^2)\\
\nonumber
\times\Big[m\Big(\frac{dt}{d\tau}\Big)\frac{d}{d\tau}(t-2a\phi)+(ma^2-r^3)\Big(\frac{d\phi}{d\tau}\Big)^2\Big]\Big\}\\
\nonumber
=\frac{2m\epsilon_0\zeta}{r^4}\Big[(a^2-2mr)\frac{dr}{d\tau}+(a^2-2mr+r^2)\frac{d}{d\tau}(t-a\phi)\Big],\\
\nonumber
\frac{d^2\phi}{d\tau^2}+\frac{1}{r^4}\Big\{ma\Big(\frac{dt}{d\tau}+\frac{dr}{d\tau}\Big)^2-2\Big[(ma^2-r^3)\frac{dr}{d\tau}+ma^2\frac{dt}{d\tau}\Big]\frac{d\phi}{d\tau}\\
\label{eom3}
+a(ma^2-r^3)\Big(\frac{d\phi}{d\tau}\Big)^2  \Big\}
=\frac{2m\epsilon_0a\zeta}{r^4}\frac{d}{d\tau}(r+t-a\phi),
\end{eqnarray}
for orbits in the equatorial plane.
In Fig. 3 we show examples of such orbits -- generated from equations (\ref{eom1})--(\ref{eom3}) -- of particles with different values for the charge-mass ratios. In the left panel repulsive electromagnetic configurations ($\zeta>0$) are considered. 
In this illustration we show that there is a critical value for the charge-mass ratio, namely $\zeta_*\simeq 0.67$, for which massive particles are driven towards the unstable Kerr retrograde photon sphere (circular red orbit).
For $\zeta<\zeta_*$, charged particles fall towards the event horizon after crossing the ergosphere. Otherwise, for $\zeta>\zeta_*$ particles are deflected towards infinite. Comparing to the case in which $\zeta=0$ (black trajectory), we see that there is an explicit deviation between orbits of neutral and charged particles due to Papapetrou fields.
In the right panel attractive configurations are considered. In this case no deviation can be detected and particles follow same trajectories towards the event horizon.

\subsection{Schwarzschild Spacetime}

Following the same standard procedure adopted in the previous section, in the case of a Schwarzschild spacetime 
in the coordinate system (\ref{sch1}) (and Killing vector $S^\mu$)
one can easily obtain the following constants of motion:
%
\begin{eqnarray}
\label{c1s}
p_t=-E,~~p_\phi=L,~~Q=p_\theta^2+L^2\cot^2\theta,    
\end{eqnarray}
so that the radial component of the canonical momenta satisfies
\begin{eqnarray}
\label{prs}
p_r=\pm\frac{\sqrt{-\Psi(r)} }{(1-2m/r)}   
\end{eqnarray}
where
\begin{eqnarray}
\nonumber
\Psi(r)=-\frac{4m^2\tilde{q}^2s_0^2-(L^2+Q+\tilde{m}^2r^2)(1-2m/r)}{r^2}\\
\label{potpsi}
+\frac{4m\tilde{q}s_0(E+\tilde{q}s_0)}{r}-(E+\tilde{q}s_0)^2.
\end{eqnarray}
From (\ref{pmu}), (\ref{ham2}), (\ref{c1s}), and (\ref{prs}) we then obtain:
\begin{eqnarray}
&&\frac{dt}{d\sigma}= E\Big(1-\frac{2m}{r}\Big)^{-1}+\tilde{q}\epsilon_0,
\\
&&\frac{dr}{d\sigma}=\pm{\sqrt{-\Psi(r)}},
\\
\label{thetas}
&&\frac{d\theta}{d\sigma}=\pm\frac{\sqrt{Q-L^2\cot^2\theta}}{r^2},
\\
\label{eqphi1}
&&\frac{d\phi}{d\sigma}=\frac{L\csc^2\theta}{r^2}.
\end{eqnarray}
From the above we see that physical orbits are those which satisfy the necessary conditions $\Psi\leq 0$ and $Q>L\cot\theta$.

In order to simplify our analysis we turn again our attention to the equatorial plane by fixing initial conditions $\theta_0=\pi/2$ and
$d\theta/d\sigma|_0=0$ (that is, $Q=0$). In this case, circular orbits with $r=r_c$ are now defined by
\begin{eqnarray}
\label{psi1}
&&\Psi(r)=0,\\
\label{psi2}
&&\frac{d\Psi}{dr}=0.    
\end{eqnarray}
It can be easily shown that the solution of (\ref{psi2}) is given by
\begin{eqnarray}
\label{rcs}
r_{c\pm}=\frac{4m^2\tilde{q}^2s_0^2-L^2\pm\sqrt{\Upsilon}}{4m\tilde{q}s_0(E+\tilde{q}s_0)-2m\tilde{m}}
\end{eqnarray}
where
\begin{eqnarray}
\Upsilon=L^4+4L^2m^2[2\tilde{q}s_0(3E+2\tilde{q}s_0)-3\tilde{m}^2]
+(4m^2\tilde{q}^2s_0^2)^2.    
\end{eqnarray}
From the above we notice that the necessary and sufficient conditions for the existence of circular orbits -- together with their stability -- are rather involved given the richness of parameters. To circumvent this issue, we consider again the fair approximation in which the Papapetrou field is sufficiently small so that terms like $\tilde{q}^2s_0^2$ may be neglected. In this case the potential (\ref{potpsi}) can be written as
\begin{eqnarray}
\label{potpsib}
\Psi(r)\simeq\frac{(L^2+\tilde{m}^2r^2)}{r^2}\Big(1-\frac{2m}{r}\Big)
+\frac{4m\tilde{q}s_0E}{r}-E^2-2\tilde{q}s_0.
\end{eqnarray}
The substitution of (\ref{potpsib}) in (\ref{psi1}) and (\ref{psi2}) then furnishes
\begin{eqnarray}
\label{lc1s}
L\simeq\pm \sqrt{\frac{m(\tilde{m}^2-2E\tilde{q}s_0)}{(r^{3/2}-3mr^{1/2})}}r^{5/4},
\end{eqnarray}
where $E$ satisfies
\begin{eqnarray}
\label{ec1s}
E\simeq(r^{3/2}-2mr^{1/2})\sqrt{\frac{\tilde{m}^2-2E\tilde{q}s_0}{r^{3/2}(r^{3/2}-3mr^{1/2})}}.
\end{eqnarray}
In this approximation, circular orbits can be found as long as
\begin{eqnarray}
\label{inqs}
\frac{\tilde{m}^2-2E\tilde{q}s_0}{r^{3/2}-3mr^{1/2}}\gtrsim 0.    
\end{eqnarray}
In the case that the LRS of (\ref{inqs})
is sufficiently small,
we obtain once again 
\begin{eqnarray}
\label{eqfnl}
\tilde{m}^2\simeq2E\tilde{q}s_0    
\end{eqnarray} 
so that massive test particles are allowed to populate the Schwarzschild photon sphere located at $r_c\simeq3m$ as long as the denominator of (\ref{inqs}) is sufficiently small.
We illustrate this behaviour in Fig. 4.
\begin{figure}
\begin{center}
\includegraphics[width=6cm,height=4cm]{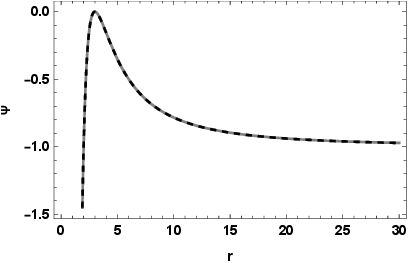}
\caption{The potential $\Psi$ according to (\ref{potpsi}). In this illustration we fixed the parameters
$Q=0$, $E=m=1$ and $L=3\sqrt{3}Em$. The grey dashed curve corresponds to the classical case in which 
$s_0=\tilde{m}=0$ so that the photonshere is located at $r_c=3m$ as one should expect. In the black dashed curve we fixed $q=0.01/s_0$ and $\tilde{m}^2\simeq2E\tilde{q}s_0\equiv0.02$. In the above we see the both curves practically coincide so that charged massive particles are able to populate the Schwarzschild photon sphere at $r_c=3m$ due to the presence of a test Papapetrou field.}
\label{fig0}
\end{center}
\end{figure}

The result obtained above might puzzle the reader once the radial component of the electric field reaches its maximum over the Schwarzschild photon sphere as mentioned by the end of Section \ref{sec:2}. However, let us consider a particle whose $4$-velocity measured by our proper observer is given by
\begin{eqnarray}
u^\mu=(\tilde{m}/E, 0, 0, 0).    
\end{eqnarray}
From (\ref{geo}) we then obtain that on the photon sphere such particle is subjected 
to a force term
\begin{eqnarray}
f^i\propto -\frac{\tilde{m}^2}{27E^2 m}\delta^i_{r}+\frac{\tilde{q}}{\tilde{m}}E^i.    
\end{eqnarray}
But
\begin{eqnarray}
E^i\equiv\frac{4s_0}{27Em}\delta^i_{r},  
\end{eqnarray}
so that
\begin{eqnarray}
f^i\propto -\frac{2}{27E^2m\tilde{m}}(\tilde{m}^2-2E\tilde{q}s_0)\simeq 0
\end{eqnarray}
according to (\ref{eqfnl}).

\section{Final Remarks}
\label{sec:4}

In this paper we examine some fundamental consequences in particle motion due to test Papapetrou fields of vacuum spacetimes. One of the underlying motivations to consider a test Papapetrou field comes from the fact that the energy-momentum tensor built with the
timelike Killing vector of Kerr metric exactly matches the electromagnetic field of a Kerr-Newman black hole, thus furnishing a physical signature 
to such Papapetrou field. Restricting ourselves to the framework of General Relativity, Kerr and Schwazschild metrics were considered. In each case, proper frames of reference were fixed.
For the Kerr spacetime we evaluate the electric and magnetic fields -- which emerge from its general Killing vector -- for a locally non-rotating (LNR) frame of reference. We show that the axial component of such vector furnishes a divergent magnetic field with no physical grounds. Taking this result into account 
we restrict ourselves to pure timelike Killing vectors.
In the case of a Schwarzschild spacetime,
a proper timelike observer was considered. In this configuration we showed that while the magnetic field vanish, the radial component of the electric field has a global maximum over the photon sphere. 
In order to probe for the effect of such electromagnetic fields -- in Kerr as in Schwarzschild spacetimes 
-- we study the motion of charged test particles in the equatorial plane.
For the case of a Schwarzschild black hole we show that massive/charged test particles may populate the 
unstable photon sphere for given domain of the parameter space.
Restricting ourselves to orbits with LNR initial conditions for the case of a Kerr black hole we show that there is an explicit deviation between orbits of neutral and charged particles in the case of repulsive electromagnetic configurations. For critical charge-mass ratios $\zeta_*$, test particles can be found in the Kerr retrograde photon sphere.

Although the numerical results presented in this paper correspond to mere mathematical simulations, it can be shown that the features presented here hold for a wide domain of parameters/initial conditions as long as 
an event horizon is formed and charged test particles are allowed to acquire proper initial conditions -- like LNR in the Kerr case. Taking into account that Papapetrou fields of Kerr spacetimes are connected to physical electromagnetic fields of Kerr-Newman black holes, the analysis presented in this paper could also shed some light on the description of more involved/realistic configurations such
as the case that electromagnetic fields interact with surrounding matter and a plasma magnetosphere is formed.
Another feature which is worth of examination is what would be the role of Papapetrou
fields in Blandford-Znajek processes \cite{Blandford:1977ds} of extracting rotational power of a black hole. 

Furthermore, in spite of the fact the physics of black holes is well established in General Relativity – at least
from a theoretical point of view at classical level – observations indicate that mass,
angular momentum and charge do not furnish a complete set of parameters to properly
describe high energy configurations as remnants black holes formed from binaries
mergers. In fact, recent data by the LIGO and Virgo collaborations \cite{virgo1,virgo2,virgo3,virgo4} established
that gravitational wave emission from binary black holes mergers were engendered from
mass ratios ranging from $\alpha\simeq 0.8$ down to $\alpha\simeq 0.5$ implying that the remnant black hole must have
a boost along a particular direction relative to an asymptotic Lorentz frame at null
infinity \cite{Aranha:2021zwf} – where such emissions have been detected. 
The extension of the results shown in this paper for the case of boosted rotating black holes
will be an object of further investigation.

\section{Acknowledgments}

RM acknowledges financial support from
FAPERJ Grant No. E-$26/010.002481/2019$.

\end{document}